\documentclass[prl,superscriptaddress,showpacs,twocolumn]{revtex4}
\usepackage{amsmath,amsfonts,amssymb}
\usepackage{graphicx}
\DeclareGraphicsExtensions{.jpg, .pdf, .tif}
\begin{document}

\title{Damage and fluctuations induce loops in optimal transport networks}

\author{Eleni Katifori}\email{ekatifori@mail.rockefeller.edu}
\affiliation{Center for Studies in Physics and Biology, The Rockefeller University}
\author{Gergely J. Sz\"oll\H{o}si}
\affiliation{Biological Physics Department, E\"otv\"os University, Budapest}
\author{Marcelo O. Magnasco}
\affiliation{Laboratory of Mathematical Physics, The Rockefeller
University, 10021 New York, NY USA}

\begin{abstract}
Leaf venation is a pervasive example of a complex biological network, endowing leaves with a transport system and mechanical resilience. Transport networks optimized for efficiency have been shown to be trees, i.e. loopless. However, dicotyledon leaf venation has a large number of closed loops, which are functional and able to transport fluid in the event of damage to any vein, including the primary veins. Inspired by leaf venation, we study two possible reasons for the existence of a high density of loops in transport networks: resilience to damage and fluctuations in load. In the first case, we seek the optimal transport network in the presence of random damage by averaging over damage to each link. In the second case, we seek the network that optimizes transport when the load is sparsely distributed: at any given time most sinks are closed. We find that both criteria lead to the presence of loops in the optimum state. 
\end{abstract}


\maketitle

\begin{figure*}[tb] 
\includegraphics[width=3.4in]{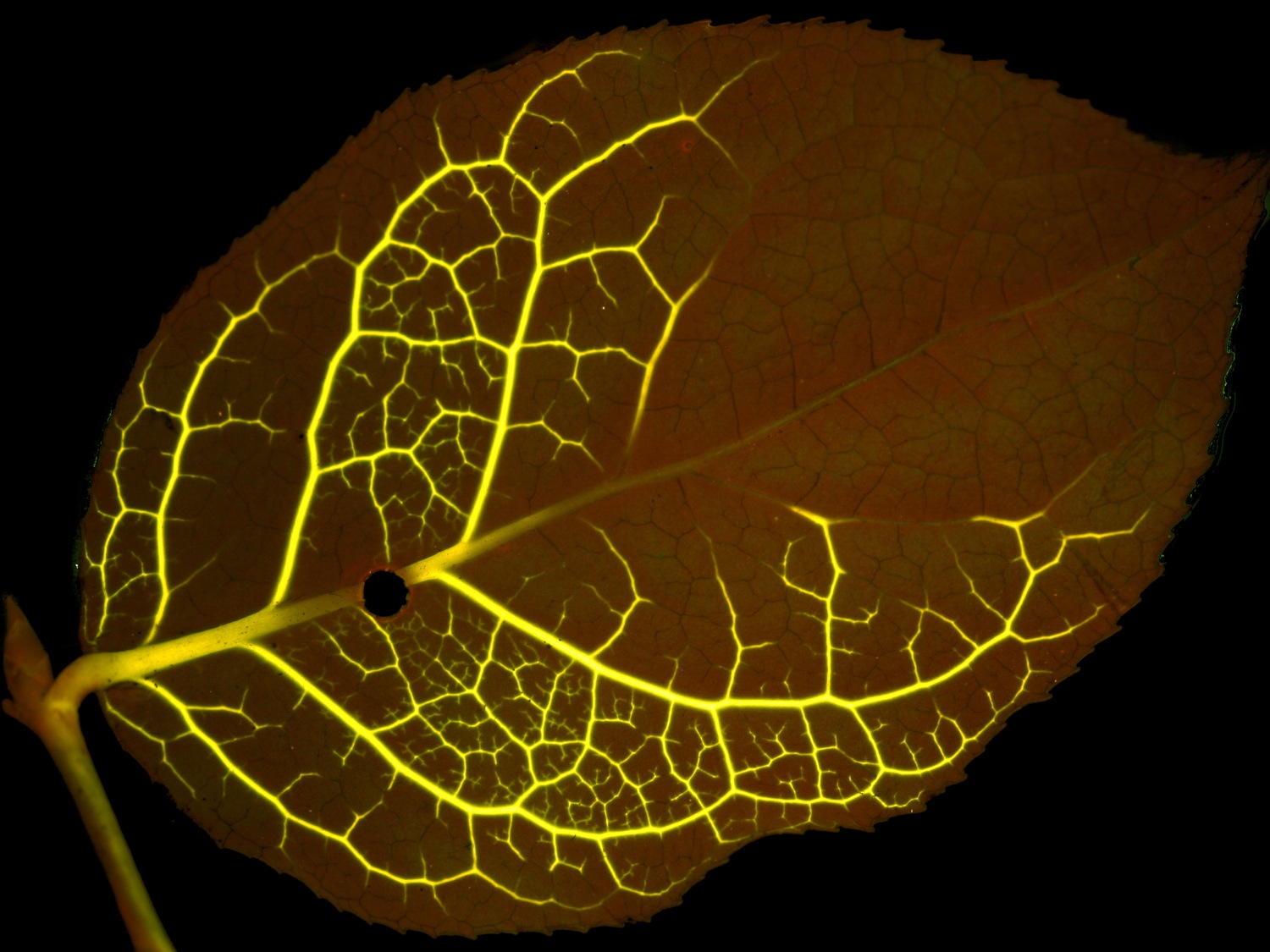}
\includegraphics[width=3.4in]{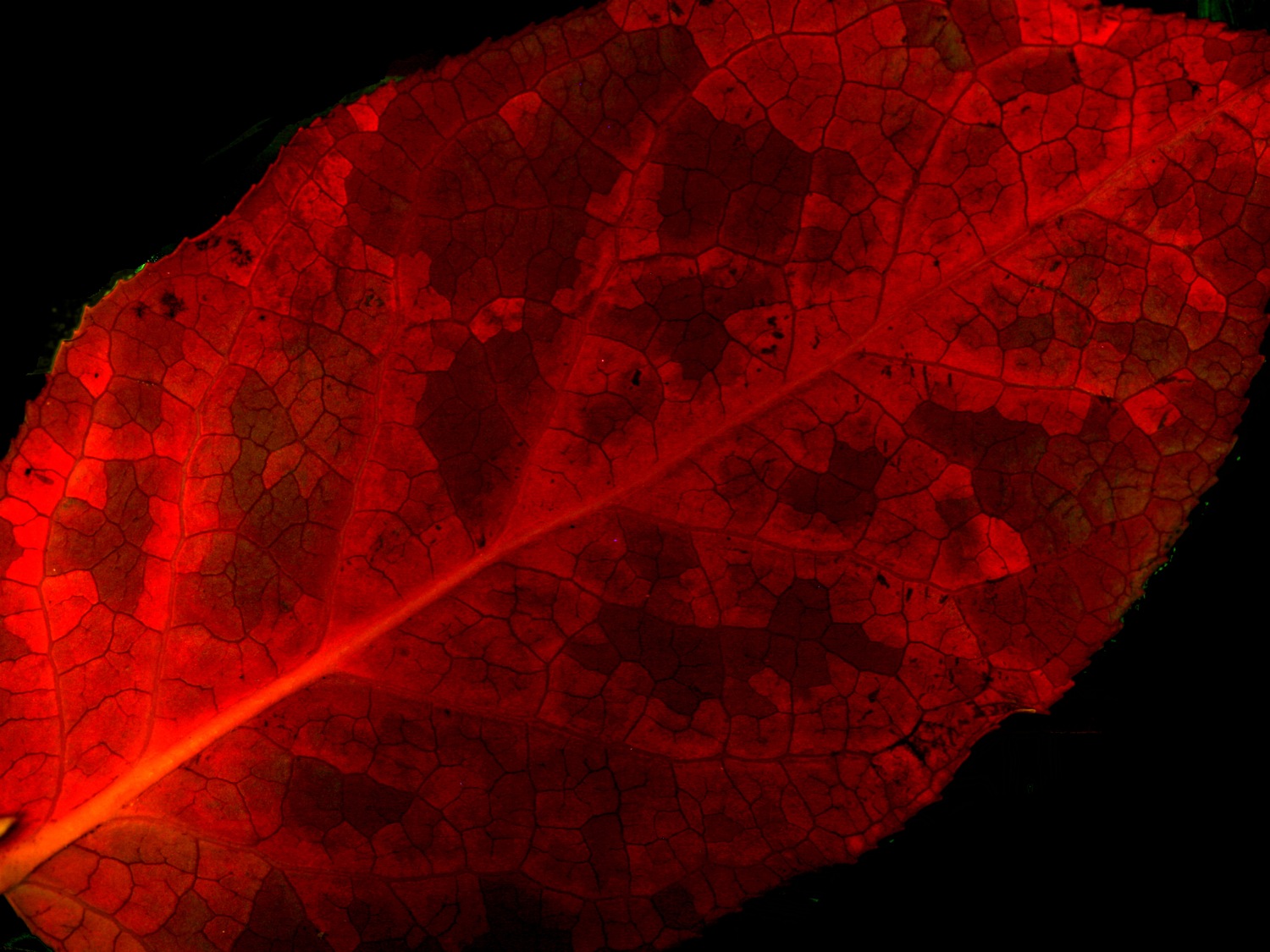}
\caption{\textbf{Left}: Flow routed around an injury in a lemon leaf.  A circular cut (black circle) was made on the main vein  of a lemon leaf; fluorescein (bright green/yellow) was injected post injury at the stem (lower left, below the field of view), and was observed to flow through the vein network around the injury, closing a number of loops and eventually reaching the tip of the leaf. The two secondary veins immediately ``downstream'' from the injury (i.e., the two attached immediately to the right of the black circle) are thus supporting reversed flow from the tertiary veins back to the primary vein. Full HD movie available as part of the Suppl. Materials. \textbf{Right}: Stomatal patchiness visualized by chlorophyll autofluorescence (455nm stimulation, broad red emission) in a lemon leaf. Entire domains of the leaf show all their stomata either closed (bright) or open (dark). 
} 
\label{fig1}
\end{figure*}

Networks optimized for various transport characteristics are of evident importance in urban infrastructure and technology; more subtly, they are also of great importance in various natural settings, such as river networks \cite{RodriguezIturbe2001}, or in biological settings where the network has been subjected to natural selection \cite{Ball2001}. The question of convexity is central in the study of such networks; for many objective functions, the optimal network can be shown to be topologically a tree, a loopless graph such that cutting any bond disconnects it into two pieces. For example, in the case of a resistor network where the cost of a conductance $C$ between two nodes is proportional to $C^\gamma$ and the total cost of the network is bounded, the network which optimizes energy dissipation will be a uniform ``sheet'' when $\gamma>1$, or a tree when $\gamma<1$, with a phase-transition-like change at $\gamma=1$ \cite{Bohn2007, Durand2007} \footnote{The parameter $\gamma$ is related to the physical nature of the network in question; i.e. assuming equal length of links, for a pipe network with Poiseuille flow $\gamma = 0.25$ whereas for a resistor network (uniform resistors, resistance only depending on thickness of wire) $\gamma = 1$.}. The networks for $\gamma<1$ are related to optimal channel networks (OCNs) \cite{RodriguezIturbe2001}, widely used to study and model river networks with large basins; the extensively studied case \cite{Bernot2008} is mostly devoted to arborescent distribution structures.  

Yet everyone is familiar with the feeling to the touch of decaying leaves on damp ground; they feel like a piece of gauze or cloth, exhibiting resistance when subject to shear forces. This would not be so were the skeleton of the leaves treelike, and in fact it evidently is not. Dicotyledon leaf venation is one of many natural networks which contain loops, in fact {\em recursively nested} sets of loops \footnote{Monocotyledon leaf venation also contains loops, although less evident. The prominent parallel veins are regularly connected with smaller perpendicular veins.}. Other examples include the vasculature of two-dimensional animal tissues such as the retina \cite{Fruttiger2002, Schaffer2006}, the structural veins of some insect wings or the structural weblike bracing of certain gorgonian corals such as sea fans, as well as the network of rivers at deltas or lowlands. Almost all artificial distribution networks contain loops, most evidently the streets in city plans (see Suppl. Materials). Animal vasculature of flat tissues, for instance retinas, shares many of the properties of leaf venation  (though mechanical stability is usually not one of its roles), while animal structural vein patterns, such as corals or insect wings, have mechanical roles unrelated to transport. As with any other natural system, we do not know a priori what (if any) is the functional being optimized in these examples, and our only starting point is to discard as a candidate any  functional or objective function whose  optimal networks differ dramatically in structure from the observed ones. Therefore we know from the outset that the functional analyzed in \cite{Bohn2007} could not possibly be the one being optimized in these cases, as for no value of $\gamma$ is the optimal network similar to the ones observed. 

Several studies have concentrated on the morphogenesis of leaf venation patterns, trying to elucidate the mechanisms by which a dense set of loops is created \cite{Nelson1997, Runions2005, Dimitrov2006, Laguna2008}. In this work we take a complementary point of view, namely, we shall want to understand why such patterns were chosen by evolution in the first place. We make  the ad hoc assumption that evolution of such patterns was not strongly constrained by choice of morphogenetic mechanism and that, were the patterns suboptimal in some fashion, some other morphogenetic algorithm would have been evolved. The interplay between morphogenetic mechanism and the optimization described herein is beyond the scope of this work. 

There are two important aspects of leaf physiology that bear directly on their survival ability. First, leaves are under constant attack, from the elements as well as pathogens, insects, and herbivores \cite{Roth-Nebelsick2001a}. Were the leaf vascular network tree-like, damage to any vein would result in the death of all the leaf section downstream from that vein, and hence entire leaf sectors would be observed to be dead. However, this is not the case, as illustrated in the experiment in Fig.~\ref{fig1}: the loops in the venation permit flow to be routed around any injury to the veins, even in case of injury to the main vein.  We thus seek to optimize leaf function {\em in the presence of damage to the veins}. 


Second, the studies that found that a tree topology optimizes transport were carried out for the case of optimization under a spatially and temporally constant load. Yet in the case of, for instance, brain vasculature, we know that the system is  sensitively optimized to deliver changing fluxes under constantly varying load; this is, in fact, the basis of the BOLD response that underlies functional magnetic resonance imaging. Indeed, there is a similar phenomenon in the case of leaves, called {\em stomatal patchiness}. Stomatal patchiness refers to heterogeneous stomatal aperture across the surface of the leaf blade (and thus water evaporation and photosynthetic activity \cite{Mott1998}). Initially considered to be an artifact or epiphenomenon, stomatal patchiness is now believed to be part of plant physiology, a possible evolutionary adaptation to situations in which ambient conditions would increase water flux beyond what it tolerable by the plant. 
A spatially and temporally irregular driving force in the form of stomatal patchiness would thus be contrary to the assumptions of the uniform load models described in Ref.~\cite{Bohn2007, Durand2007}. Because of the nonlinearity of the optimization cost function, spatio-temporal irregularities can potentially alter the optima. In particular, the proofs of looplessness in Refs.~\cite{Banavar2000, Bohn2007, Durand2007} all rely rather directly and essentially on constant fluxes. Therefore we also seek to optimize leaf function under conditions of varying load. 

\begin{figure}[tb] 
\includegraphics[width=3.4in]{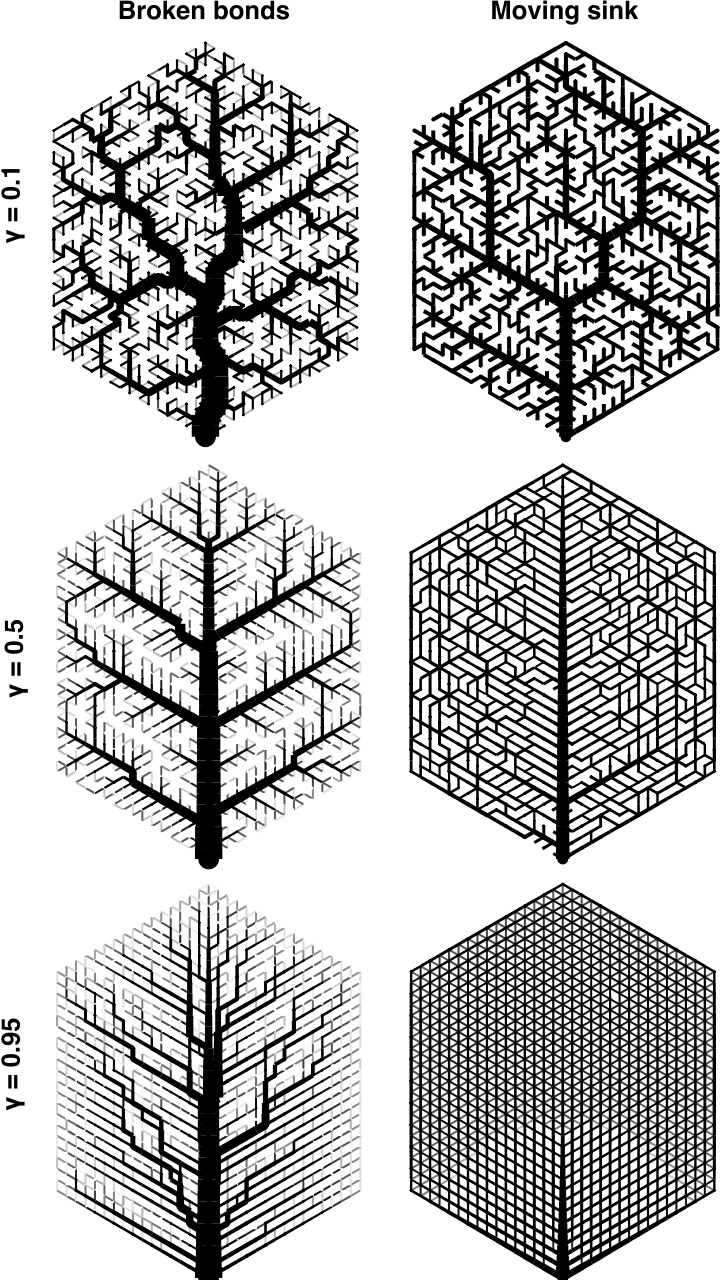}
\caption{Loops as a result of optimizing under damage to links (left column) and under a fluctuating load (right column). In all plots the vein thickness is proportional to $C^{(\gamma+1/2)/3}$.  }
\label{fig2}
\end{figure}

Optimization under damage to veins implies, virtually by definition, the formation of loops. Even then, we shall find that more loops are formed than the minimum strictly required by topology, which is a ring joining all outermost nodes. More surprisingly, we shall find that optimization under a varying load leads to the formation of dense, recursively-looped structures. 

We now define the two models more precisely. We consider, following \cite{Bohn2007, Durand2007}, a network consisting of nodes $k$, joined by conductances $C_{kj}$; we shall denote by $\langle j,k\rangle$ the set of nodes $j$ adjacent to a given node $k$. A conductance $C$ is assumed to ``cost'' an amount $C^{\gamma}$, and the total cost of the network to be constant and equal to 1. 
\begin{equation} \frac{1}{2}\sum_k{  \sum_{\langle j,k\rangle} C_{kj}^\gamma } = 1\end{equation}
At any given node $k$ there are currents $I_{kj}$ through the edges of the graph, whose sum are net currents $\mathcal{I}_i$ which correspond either to evaporation through stomata or water injection at the stem. The currents are driven by differences in the potential $V_i$ at the nodes, as $I_{kj}=C_{kj}(V_k-V_j)$: 
\begin{equation} \sum _{\langle j,k\rangle} C_{kj} (V_k - V_j) = \mathcal{I}_k \end{equation}
an equation whose inversion yields the $V_k$ and consequently the $I_{kj}$. Finally, the functional being optimized is the total power dissipation 
\begin{equation} P =\frac{1}{2} \sum_k{  \sum_{\langle j,k\rangle} C_{kj} (V_k-V_j)^2}.  \end{equation}
These equations need to be supplemented with the stipulation that $\mathcal{I}_1 = N-1$ (where $k=1$ is the stem) and $\mathcal{I}_k = -1$ for $k>1$. 

In the first model (robustness to damage), we compute the power dissipated when one of the links (e.g. $ab$) is broken: $C^{ab}_{kj} = C_{kj} (1-\delta_{ak}\delta_{bj}-\delta_{aj}\delta_{bk})$. For each set of conductivities with a cut bond we can compute $V_k^{ab}$ and subsequently obtain the dissipation $P^{ab}$. The total dissipation is defined as:
\begin{equation} 
R = \sum_{(ab)} P^{ab}\end{equation} 
which is the functional being minimized.
Note that if breaking $(ab)$ disconnects the graph, then $P^{ab}$ becomes infinite. Thus, finiteness of $R$ requires that breaking any one bond should not disconnect the graph. This can be satisfied by adding a perimeter ring of conductances (no matter how small) to the terminal nodes of any tree. If the tree had $T$ terminal nodes, then this would result in $T$ facets in the graph. Many leaves have such a rim of veins \cite{Roth-Nebelsick2001a}. However this only guarantees finiteness; to actually minimize the value of $R$ more loops than that are required, as shown in the right column of Fig.~\ref{fig2}. 


In our second model, we consider fluctuating load by introducing a single moving sink. We define $\mathcal{I}^{a}_k = \delta_{0k} - \delta_{ak}$, i.e., a single source at the stem and a single sink at node $a$. This induces new potentials $V^a_k$, power loss $P^a$ and similarly to the cut bond model:
\begin{equation}
F = \sum_a P^a. \end{equation}
Optima of this functional are shown in Fig.~\ref{fig2}, right column. 

\begin{figure}[tb] 
\includegraphics[width=3.4in]{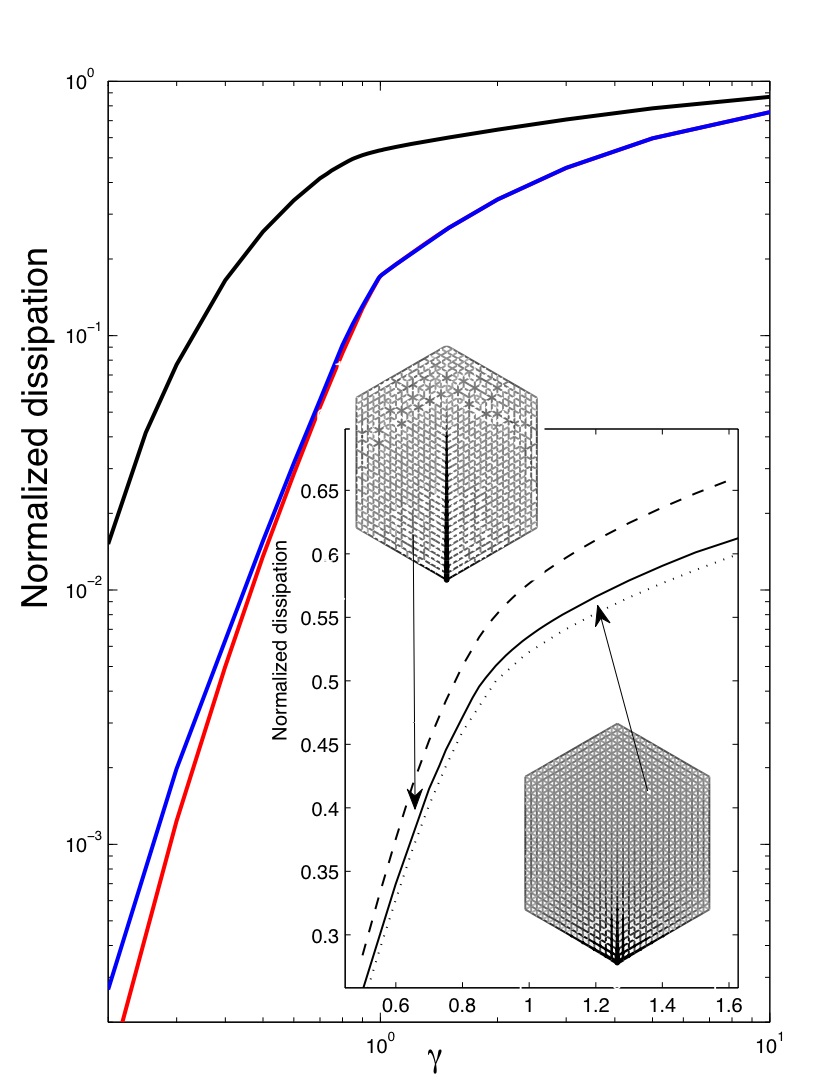}
\caption{Normalized network dissipation vs $\gamma$. Red line: homogeneous load, blue line: cut bond, black line: fluctuating sink. The number of network nodes  was $N=949$. Inset: Close up of the crossover for the fluctuating sink case. Solid line: medium size network ($N=949$), dashed line: small network ($N=285$), dotted line: big network ($N=1589$). }
\label{fig3}
\end{figure}

The model in \cite{Bohn2007, Durand2007, Banavar2000} was already computationally costly enough that it could only be optimized through relaxation techniques, or Monte Carlo methods that involved searching the space of tree graphs. The extensions presented in this paper cannot be optimized by such a Monte Carlo method as the optima are not trees.  We alternatively use an annealing method, described in the Supplementary Material. However, much like any method that does not use exhaustive search, the minima found cannot be guaranteed to be global minima. The method required the evaluation of the inverse of an $N \times N$ matrix $\hat{G}$ that depends on the conductivity. For computational efficiency in the cut bond model, rather than inverting a different matrix for each of the terms in the $ab$ sum, we perturbatively expand $\hat{G}$ and show that it can be evaluated using only the inverse of the original matrix of the full network and a sparse matrix multiplication, dramatically speeding up the evaluation. To simplify the calculation in the fluctuating load model we similarly use the sparse nature of the $\mathcal{I}^a$. Details are given in Supplementary Material.

Similarly to the results found in \cite{Bohn2007, Durand2007, Banavar2000} the cut bond model exhibits a transition from the $\gamma<1$ hierarchical structure case, to the $\gamma>1$ case, where the network is a uniform sheet. However, unlike the simple tree model, the veins anastomose and we observe the formation of nested loops, reminiscent of the ones seen in real leaves. Indeed, evolutionary trends in vascular plants indicate a tendency of the vascular system to develop redundancy and hierarchical network patterns \cite{Roth-Nebelsick2001a}.

As shown in Fig.~\ref{fig3}, where we plot the ratio of the energy of the network to the energy of a network with a constant conductivity distribution and same $\gamma$ (normalized dissipation), the results for the bond case are very similar to the uniform load case. They both exhibit a cusp at $\gamma=1$. The situation is drastically different for the moving sink model, where the phase transition has been replaced by a crossover shown in the inset of Fig.~\ref{fig3}.

Although in the sink model we still observe hierarchical loops, the quantitative aspects of the the networks in the two cases are very different. For example, the average node valency is smaller for the sink model for $\gamma<~0.4$, and becomes larger after that. In both cases, for $\gamma>1$ it asymptotically approaches 6. Moreover, whereas for the broken bond model fractal-like features of the network persist until $\gamma<~1$, for the sink model fractality disappears for $\gamma>~0.5$ \cite{Kull1995}. 

To conclude, it is widely yet incorrectly stated that many natural distribution networks, such as animal vasculature or tree leaves, are treelike \cite{Pelletier2000,Xia2007}; even the most cursory of visual inspections rapidly refutes such statements. Two fundamental misconceptions underlie these statements. First, an unstated yet universal implication is that the trees are hierarchically ordered, i.e., as they are inevitably depicted as regular binary trees, one gets the impression of an ordered branching from the top level, first order to second and subsequent orders. Second, the venous and arterial tree are described as touching each other strictly at the level of the ``terminal nodes'' of each tree, the capillaries, by implication the only member of the hierarchy where it is ambiguous whether it's a vein or an artery. This description is disproved by looking at real leaves or images of, for instance, retina, where it is evident that venous capillaries of all orders impinge directly on primary arteries and viceversa, and that, moreover, veins of any order branch directly from higher-order veins of much larger order \cite{Fruttiger2002, Schaffer2006}. This topological disorder permits circulation in case of obstacles, but may also, as our results above suggest, confer superior deliverance to fluctuating loads.

We're deeply indebted to Raymond Raad, Steffen Bohn, and Yves Couder; we'd like to acknowledge help from Keith Mott. During preparation of this manuscript we became aware of the work of Francis Corson.

\bibliography{LeafPaper}

\end{document}